\documentstyle[12pt]{article}
\newcommand{\PSbox}[3]{\mbox{\rule{0in}{#3}\includegraphics{#1}\hspace{#2}}}

\newcommand{\mysection}[1]{\setcounter{equation}{0}\section{#1}}

\newcommand{\nc}{\newcommand}
\nc{\beq}{\begin{equation}} \nc{\eeq}{\end{equation}}
\nc{\beqa}{\begin{eqnarray}} \nc{\eeqa}{\end{eqnarray}}
\nc{\lsim}{\begin{array}{c}\,\sim\vspace{-21pt}\\< \end{array}}
\nc{\gsim}{\begin{array}{c}\sim\vspace{-21pt}\\> \end{array}}
\newfont\figfont{cmr7 scaled 1200}
\begin{document}

\begin{titlepage}

\begin{center}
{\hbox to\hsize{hep-ph/9606414 \hfill  MIT-CTP-2542}}

\bigskip

\bigskip

{\Large \bf  The Minimal Supersymmetric Standard Model (MSSM) 
} \\

\bigskip

\bigskip

{\bf Csaba Cs\'aki \footnote{Supported in part by the
DOE under cooperative agreement \# DE-FC02-94ER40818.}}\\

\smallskip

{ \small \it Center for Theoretical Physics

Laboratory for Nuclear Science and Department of Physics

Massachusetts Institute of Technology

Cambridge, MA 02139, USA }

 \bigskip

\vspace{2cm}

{\bf Abstract}\\[-0.05in]
\end{center}

The structure of the MSSM is reviewed. We first motivate the 
particle content of the theory by examining the quantum numbers 
of the known standard model particles and by the 
requirement of anomaly cancellation. 

Once the particle content is fixed we can write 
down the most general 
renormalizable superpotential. However such a
superpotential will contain terms breaking lepton and baryon 
number  which leads us to the
concept of R-parity conservation.

The question of supersymmetry breaking is discussed next.
We list the possible soft breaking terms. However the 
Lagrangian involving the most general soft breaking terms
is phenomenologically intractable because of the appearance of
the many new parameters. It also leads to some unacceptable 
predictions. To reduce the number of parameters we
restrict ourself to the case with universal soft breaking terms
at the GUT scale.  
We motivate the need for universal soft breaking terms 
by the apparent unification of gauge 
couplings in the MSSM and by the absence of flavor changing
neutral currents. Then we discuss radiative electroweak
symmetry breaking. Radiative breaking arises because 
the one loop corrections involving
the large top Yukawa coupling change the sign of the 
soft breaking mass parameter of the up-type Higgs doublet, 
this way introducing a nontrivial minimum in the Higgs
potential. 

Finally we give an overview of the possible mixings in the 
MSSM and enumerate the physical (mass eigenstate) fields 
together with the mass matrices.

\end{titlepage}

\mysection{Introduction}

The Standard Model (SM) of particle physics enjoys an 
unprecedented success: up to now no single experiment
has been able to produce results contradicting this model.
Particle theorists are nevertheless unhappy with this theory.
The most important features of the SM that are technically 
allowed but nevertheless theoretically unsatisfying are the 
following:

a. There are too many free parameters

b. The SU(2)$\otimes $U(1) group is not asymptotically free

c. Electric charge is not quantized

d. The hierarchy problem.

\noindent While the first three problems can be taken care by 
introducing grand unification, the mystery of the hierarchy
problem remains unsolved in GUT's as well. The hierarchy problem
is associated with the presence of elementary scalars (Higgs)
in the SM. The problem is that in a general QFT containing
an elementary scalar the mass of this scalar would be naturally
at the scale of the cutoff of the theory (if the SM were the
full story then the Higgs mass would be naturally of
${\cal O} (M_{Pl})$) due to the quadratically divergent
loop corrections to the Higgs mass. If one wants to protect
the scalar masses from getting these large loop corrections
one needs to introduce a new symmetry. The only known such
symmetry is supersymmetry (SUSY), which relates fermions 
and bosons to each other.\footnote{Another way of solving
the hierarchy problem is to assume that the Higgs is not an 
elementary scalar but a bound state of fermions, which
idea leads to technicolor theories.}

In this paper we will review the minimal extension of the SM
that includes SUSY, the Minimal Supersymmetric Standard 
Model (MSSM) \cite{Haber,Nilles,Baer,Bagger,Diego,Martin}. 
We will assume that the reader is familiar 
with both the structure of the SM and with N=1 global SUSY.

The paper is organized as follows: Section 2 discusses the 
particle content of the MSSM together with the most general
renormalizable superpotential. The need for R-parity is also
shown here. The possible form of SUSY breaking terms 
is described in Section 3 together with the phenomenological
constraints (and theoretical biases) on these SUSY breaking
terms. That section will be closed with a discussion of radiative
electroweak symmetry breaking. In Section 4 we present the
possible mixings in the MSSM together with the mass matrices.
Finally we conclude in Section 5. We did not attempt to give
a comprehensive list of references of the subject. More
complete references on the MSSM can be found in 
\cite{Haber,Nilles,Baer,Bagger,Martin}.

\mysection{Particle content and superpotential}

The SM is a spontaneously broken 
SU(3)$\otimes $SU(2)$\otimes $U(1) gauge theory with the matter 
fields being

\begin{eqnarray}
\mbox{leptons:}\; \;  L_i =  \left( \begin{array}{c} 
\nu \\ e \end{array}
\right)_{L_i} = &(1,2,-\frac{1}{2})& \nonumber \\
e_{R_i}=&(1,1,-1)& \nonumber \\
\mbox{quarks:}\; \;  Q_i= \left( \begin{array}{c} 
u \\ d \end{array}
\right)_{L_i} = &(3,2,\frac{1}{6})& \nonumber \\
u_{R_i}=&(3,1,\frac{2}{3})& \nonumber \\
d_{R_i}=&(3,1,-\frac{1}{3})& \nonumber \\
\mbox{Higgs:}\; \;  H= \left( \begin{array}{c} 
h^+ \\ h^0 \end{array}
\right) = &(1,2,\frac{1}{2})& \; i=1,2,3, 
\end{eqnarray}
where $i$ is the generation index, $L$ and $R$ refer to
left and right handed components of fermions and the numbers
in parenthesis are the SU(3)$\otimes $SU(2)$\otimes $U(1)
quantum numbers.

The MSSM is an extension of the SU(3)$\otimes $SU(2)$\otimes $
U(1) gauge theory with N=1 SUSY (which will be broken in a 
specific way, see Section 3).

The rules of building N=1 SUSY gauge theories are to assign
a vector superfield (VSF) to each gauge field and a chiral
superfield ($\chi$SF) to each matter field. The physical 
particle content of a VSF is one gauge boson and a Weyl fermion
called gaugino, and of the $\chi$SF is one Weyl fermion
and one complex scalar \cite{WB,Ross}. 
The VSF's transform under the adjoint
of the gauge group while the $\chi$SF's can be in any 
representation. Since none of the matter fermions of the
SM transform under the adjoint of the gauge group we can not
identify them with the gauginos. Thus we have to introduce 
new fermionic SUSY partners to each SM gauge boson.

If we now look at the matter fields of the SM listed above 
we see that the only possibility to have two SM fields 
as each others superpartner would be to have
$\tilde{H}=i\tau_2 H^*$ as a superpartner of L. However
this is phenomenologically unacceptable since L carries
lepton number 1, while H lepton number 0, and the superpartners
must carry the same gauge and global quantum numbers. Thus we
conclude that we have to introduce a new superpartner field
to every single field present in the SM: scalar partners 
to the fermions (called sleptons and squarks), fermionic 
partners to the Higgs (Higgsino) and gauge bosons (gaugino).

However we can see that we have introduced one extra
fermionic SU(2) doublet Higgs 
with SU(3)$\otimes $ SU(2)$\otimes $U(1)
quantum numbers (1,2,$\frac{1}{2}$). This is unacceptable
because of the Witten anomaly and because of the U(1) anomaly
that it causes. Thus we need to introduce one more SU(2)
doublet with opposite U(1) charge. The need for this
second Higgs doublet can also be seen in a different way:
in the SM one needed only one Higgs doublet to give masses
to both up and down type quarks, because one was
able to use both $H$ and $\tilde{H} $ in the Lagrangian. However
in SUSY theories the superpotential (the only source of
Yukawa interactions between only matter fields and its partners)
must be a holomorphic function of the fields thus both
$H$ and $\tilde{H}$ can not appear at the same time in the
superpotential \cite{WB}. 
This again calls for the need of two Higgs
doublet $\chi$SF's, one with the quantum number of $H$, the
other with the quantum numbers of $\tilde{H}$. The final 
resulting $\chi$SF content of the MSSM is given in table 1. 
(We use the conjugate fields $\bar{U},\bar{D},\bar{E}$ 
because in the superpotential we can not use conjugation 
anymore.) 

\begin{table}
\begin{center}

\begin{tabular}{rrrrrr}
$\chi$SF & SU(3) & SU(2) & U(1) & B & L \\
$L_i$ & 1 & 2 & $-\frac{1}{2}$ & 0 & 1 \\
$\bar{E}_i $ & 1 & 1 & 1 & 0 & $-1$ \\
$Q_i$ & 3 & 2 & $\frac{1}{6}$ & $\frac{1}{3}$ & 0 \\
$\bar{U}_i$ & $\bar{3}$ & 1 & $-\frac{2}{3}$ & $-\frac{1}{3}$ & 
0 \\
$\bar{D}_i$ & $\bar{3}$
 & 1 & $\frac{1}{3}$ & $-\frac{1}{3}$ & 0 \\
$H_1$ & 1 & 2 & $-\frac{1}{2}$ & 0 & 0 \\
$H_2$ & 1 & 2 & $\frac{1}{2}$ & 0 & 0 \end{tabular}
\end{center}

\caption{The $\chi$SF's of the MSSM with their gauge and global
quantum numbers. $i=1,2,3.$}
\end{table}

Once the particle content is fixed one can try to write down 
the most general renormalizable Lagrangian for this 
N=1 SUSY SU(3)$\otimes $SU(2)$\otimes $U(1) theory. It is known
from the structure of N=1 SUSY gauge theories that the
Lagrangian is completely fixed by gauge invariance and by
supersymmetry, except for the choice of the superpotential,
which could contain all possible gauge invariant operators 
of dimensions not greater than 3. In our case this means 
that 

\begin{eqnarray}
W&=&(\lambda_u^{ij} Q^iH_2\bar{U}^j+\lambda_d^{ij}Q^iH_1\bar{D}^j
+\lambda_e^{ij}L^iH_1\bar{E}^j+\mu H_1H_2) + \nonumber \\
 &+&(\alpha_1^{ijk} Q^iL^j\bar{D}^k+
\alpha_2^{ijk}L^iL^j\bar{E}^k
+\alpha_3^iL^iH_2+\alpha_4^{ijk}\bar{D}^i\bar{D}^j\bar{U}^k).
\end{eqnarray}
The terms in the first pair of parenthesis correspond to 
the SUSY extension of the ordinary Yukawa interactions of
the SM and an additional term (``$\mu$-term'') breaking the
Peccei-Quinn symmetry of the two doublet model. However the
terms is the second pair of parenthesis break baryon and lepton
number conservation. Thus as opposed to the SM where the
most general renormalizable gauge invariant Lagrangian 
automatically conserved baryon and lepton number, here one has
to require some additional symmetries to get rid of the
B and L violating interactions that are phenomenologically
unacceptable. 

The easiest way to achieve this is to introduce R-parity and 
require R-parity conservation.\footnote{One could forbid 
the appearance of the B,L breaking terms by imposing different
symmetry requirements. For example a $Z_2$ subgroup of 
B$\otimes $L known as matter parity could achieve this 
goal as well. The point is that once those terms are absent
R-parity will necessarily be a symmetry of the Lagrangian.}
 Under R-parity
\begin{eqnarray}
\label{rparity}
& & H_1, H_2 \rightarrow  H_1,H_2, \nonumber \\
& & Q,\bar{U},\bar{D},L,\bar{E} \rightarrow  
-(Q,\bar{U},\bar{D},L,\bar{E}) \nonumber \\
& & \theta \rightarrow -\theta,
\end{eqnarray}
which means that 
\begin{eqnarray}
(\mbox{ordinary particle}) &\rightarrow &(\mbox{ordinary 
particle}) \nonumber \\
(\mbox{superpartner}) &\rightarrow &-(\mbox{superpartner}).
\end{eqnarray}
 Note that this $Z_2$ group is a subgroup of a U(1)$_R$ symmetry
where the R-charges of the $\chi$SF's are:
\begin{eqnarray}
& & R=1 \; \; \mbox{for} \; \; H_1,H_2 \nonumber \\
& & R=\frac{1}{2}\; \;  
\mbox{for} \; \; L,\bar{E},Q,\bar{U},\bar{D}.
\end{eqnarray}
However the imposition of the full U(1)$_R$ symmetry 
forbids Majorana masses for the
gauginos which are phenomenologically needed. There are two
possible solutions to this problem. One could impose
only 
the $Z_2$ subgroup, R-parity, which forbids the B,L violating
terms in the superpotential, but allows for gaugino mass 
terms.\footnote{R-parity as defined in eq. 2.3 is actually 
$Z_2$ subgroup of the continous R-symmetry of 2.5 combined
with a baryon and lepton number transformation. The value of 
the R-parity can be given by $R=(-1)^{3B+L+2S}$, where B is 
the baryon number, L the lepton number and S the spin of a 
given particle.} If however one imposes the full U(1)$_R$
symmetry then this symmetry has to be spontaneously broken
to its $Z_2$ subgroup leading to complications with the
resulting Goldstone boson. We will not discuss this 
possibility further here. In both cases however R-parity
is an unbroken symmetry of the theory.

As a consequence of R-parity conservation
superpartners can be produced only
in pairs, implying that the lightest superpartner (LSP) is 
stable if R-parity is exact. Most of the experimental detection
modes of SUSY are based on this fact \cite{Exp}.

\mysection{SUSY breaking, radiative breaking of 
SU(2)$\otimes $U(1)}

In the previous section we have seen the particle content and
the superpotential of the MSSM. However we know that this can 
not be the full story for two reasons:

-SUSY is not yet broken

-SU(2)$\otimes $U(1) is not yet broken.

\noindent First we discuss 
SUSY breaking. SUSY was invented to solve the
hierarchy problem. However SUSY can not be an exact symmetry
of nature since in this case many of the superpartners
should have been observed by experiments. 
One has two possibilities
for SUSY breaking, either explicit or spontaneous
 breaking. While theoretically spontaneous breaking of SUSY
is much more appealing, one nevertheless has to rule out this
possibility in the context of MSSM. To see the reason 
behind this
we have to examine the scalar quark mass matrix in detail
\cite{DimGeor}. The most general scalar mass matrix in N=1 SUSY 
gauge theories is given by \cite{Ross}
\begin{equation}
\label{scalarmass}
{M^2}^a_c=\left[ \begin{array}{cc} \bar{W}^{ab}W_{bc} +
\frac{1}{2} D^a_{\alpha} D_{\alpha c} +\frac{1}{2} D^a_{c \alpha}
D_{\alpha} & \bar{W}^{abc}W_b +\frac{1}{2}D^a_{\alpha}
D^c_{\alpha} \\ \bar{W}_{abc}^b +\frac{1}{2}D_{\alpha a}
D_{\alpha c} & W_{ab}\bar{W}^{bc} +\frac{1}{2}D_{\alpha a}
D_{\alpha}^c +\frac{1}{2} D^c_{\alpha a}D_{\alpha} \end{array}
\right] ,
\end{equation}
where $W_a=\frac{\partial W}{\partial \phi_a}|_{\phi =\langle 
\phi \rangle}$, $W_{ab}=\frac{\partial^2 W}{\partial \phi_a 
\partial \phi_b}|_{\phi =\langle 
\phi \rangle}$, etc. and $D_{\alpha}=g_{\alpha} 
\phi^{\dagger a} {T_{\alpha}}^a_b \phi_b$, $D_{\alpha c}=
\frac{\partial D_{\alpha}}{\partial \phi_c}|_{\phi =\langle 
\phi \rangle}$, etc., $W$ is the superpotential, the 
$\phi_a$'s are
the complex scalars of the $\chi$SF's, the $g_{\alpha}$'s are
the gauge couplings, and the ${T_{\alpha}}^a_b$'s are the
generators of the gauge group in the representations of the
$\chi$SF's. 

Specifying this matrix to the squarks  we note that since all
the squark VEV's must vanish (so as color and electric charge
are unbroken symmetries) $ D^a_{\alpha}=0$ for the squarks. 
On the other hand quarks get their masses solely from the
superpotential thus $\bar{W}^{ab}W_{bc}$ is nothing but the
square of the quark mass matrix $m$. Since electric charge 
and color are not broken one needs to have $D_1=D_2=0$ (where 
1 and
2 here are SU(2) indices) and $D_i=0$ ($i=1,...8$ of SU(3)). 
Thus the
only possible non-vanishing D-terms are $D_3$ and $D_Y$. 
Therefore the 
squark mass matrices can be written in the form 
\begin{equation}
\label{up}
M^2_{2/3}=\left[ \begin{array}{cc} m_{2/3}m^{\dagger}_{2/3} +
(\frac{1}{2}gD_3+\frac{1}{6}g'D_Y)1 & \Delta \\
\Delta^{\dagger} & m^{\dagger}_{2/3}m_{2/3} -\frac{2}{3} g'D_Y 1
\end{array} \right] 
\end{equation}
for the charge 2/3 squarks and
\begin{equation}
\label{down}
M^2_{1/3}=\left[ \begin{array}{cc} m_{1/3}m^{\dagger}_{1/3} +
(-\frac{1}{2}gD_3+\frac{1}{6}g'D_Y)1 & \Delta ' \\
\Delta'^{\dagger} & m^{\dagger}_{1/3}m_{1/3} +\frac{1}{3} g'D_Y 1
\end{array} \right] 
\end{equation}
for the charge -1/3 squarks. Here $m_{1/3}$ and
$m_{2/3}$ are the 3 by 3 quark mass matrices in generation space.
$M_{2/3}^2$ and $M_{1/3}^2$ are thus 6 by 6 matrices for the
3 generations of left and right handed squarks. 
 The exact form
 of $\Delta$ and $\Delta '$ is not
important for us. One may notice that 
(as a consequence of the tracelessness
of the group generators) the sum of the D-terms appearing in 
the two 
squark matrices is zero. Therefore at least one of the 
appearing D-terms
is non-positive. Assume for example that 
$\frac{1}{2}gD_3+\frac{1}{6}g'D_Y
\leq 0$. But if $\beta $ is the normalized eigenvector of the
quark mass matrix $m_{2/3}$ corresponding to the smallest 
eigenvalue
$m_0$ we get that
\begin{equation}
(\beta^{\dagger},0) M^2_{2/3} \left( \begin{array}{c} 
\beta \\ 0 \end{array}
\right) \leq m_0^2.
\end{equation}
Therefore there must be a charged scalar state with mass 
less than 
the mass of either the u or d the quark which is 
experimentally excluded. 

Thus we conclude that we need to introduce explicit SUSY breaking
terms in order to circumvent the previous argument.
However these terms must be such that the solution of the
hierarchy problem is not spoiled. Such terms are called 
soft SUSY breaking terms, and those are the terms
that do not reintroduce quadratic divergences into the theory.

The philosophy behind these soft breaking terms is the following:
there is a sector of physics that breaks SUSY spontaneously.
This is at much higher energy scales than the weak scale. 
SUSY breaking is communicated in some way (either through
gauge interactions or through gravity) to the MSSM fields and
as a result the soft breaking terms appear. One popular 
implementation
of this idea is to break SUSY spontaneously in a ``hidden 
sector'', that
is in a sector of fields that do not interact with the SM 
particles
(``visible sector'') except through supergravity which will 
mediate 
the SUSY breaking terms to the visible sector. This mechanism 
with 
minimal supergravity generates universal soft breaking terms 
for the visible
sector fields at the Planck scale. 

Thus one has to handle the MSSM as an effective theory, valid
below a certain scale (of new physics), and the soft breaking
terms will parametrize our ignorance of the details of the
physics of the SUSY breaking sector.

The most general soft SUSY breaking terms are \cite{GG}

i. gaugino mass terms

ii. scalar mass terms

iii. scalar quadratic and trilinear interaction terms.

Thus if one wants to implement this program consistently one
has to add a separate mass term for each scalar and gaugino
and add each quadratic and trilinear interaction term 
appearing in the superpotential with different coefficients
to the Lagrangian:

\begin{eqnarray}
 & & -{\cal L}_{soft}=\sum_{i=Q_{i},\bar{U}_{i},...} m_i^2 
|\phi_i|^2+ \left( 
\sum_{i=1,2,3} M_i \lambda_i \lambda_i - B\mu H_1H_2+
\right.
\nonumber \\
& & +\left. \sum_{ij} A_u^{ij}\lambda_u^{ij} Q^iH_2\bar{U}^j+
\sum_{ij} A_d^{ij}\lambda_d^{ij}Q^iH_1\bar{D}^j
+\sum_{ij} A_e^{ij}\lambda_e^{ij}L^iH_1\bar{E}^j + h. c. \right)
\end{eqnarray}
This would mean that we introduce 17 new real and 31 new complex
parameters into the theory. There are two major problems 
with this:

-not every set of $(m_i,M_i,B,A_k^{ij})$ parameters is allowed
by phenomenology

-there are too many new parameters to handle the phenomenology.

Let's first see what the requirements for the soft breaking 
parameters are. The two most serious restrictions come from
the requirements that

1. large flavor changing neutral currents (FCNC) and 
lepton number violations are absent

2. the theory should not yield too large CP violation.

One can easily understand why a general set of soft breaking 
parameters introduces large FCNC's. Let's look at the 
$K^0-\bar{K}^0$ mixing. In the SM one gets contributions from 
the diagrams shown in Fig. 1.
However in the MSSM one has additional contributions from the 
diagrams of Fig. 2, where the intermediate lines are now 
gauginos and
squarks, and the cross denotes the soft breaking squark masses.
In Fig. 2. the 
usual CKM factors appear at the vertices. Thus the leading part
of this diagram is proportional to $V^{\dagger} M^2 V$, where
$V$ is the CKM matrix. The succesfull implementation
of the GIM mechanism in the SM in $K^0-\bar{K}^0$ mixing 
is based on the fact that the diagrams are proportional
to $V^{\dagger}V=1$. However if $M^2$ is an arbitrary matrix
then $V^{\dagger}M^2V\neq 1$. Thus we can see that in order to
 maintain the successfull GIM prediction in the MSSM one has to
require that $M^2\approx m^2 1$, that is squarks must be nearly
 degenerate.

\begin{figure}
\begin{center}
\PSbox{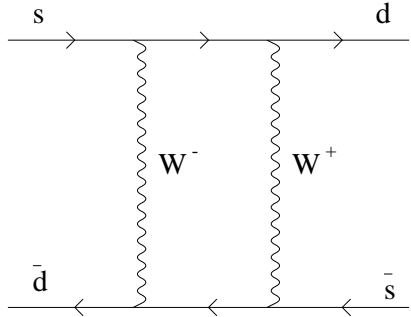 hoffset=-125 voffset=-210
hscale=70 vscale=70}{5cm}{5cm}
\end{center}
\caption{Diagrams contributing to $K^0-\bar{K}^0$ mixing 
in the SM.}
\end{figure}

\begin{figure}
\begin{center}
\PSbox{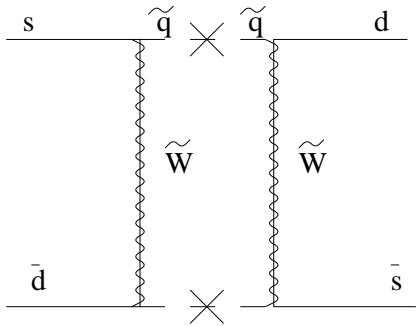 hoffset=-125 voffset=-210
hscale=70 vscale=70}{5cm}{5cm}
\end{center}
\caption{Additional diagrams contributing to $K^0-\bar{K}^0$
 mixing in the
MSSM.}
\end{figure}

Very similar arguments hold for the $\mu \to e\gamma$ process
which will result in the need of nearly degenerate sleptons.

The second constraint on the soft breaking terms comes from the
fact that the SM can account for all the measured CP violation.
Thus there is no need for extra sources of CP violation
in the MSSM, therefore it is usually assumed that the
soft breaking parameters are real. 

Thus we have seen what the phenomenological constraints
on these soft breaking parameters are. Now we present a set
of assumptions that satisfy these constraints and at the
same time highly reduce the number of free parameters of the
model:

1. Gaugino unification (common mass for the gauginos at the 
Planck scale)

2. Unification of soft masses (common soft breaking mass terms 
for the scalars
at the Planck scale)

3. Unification of the soft breaking trilinear coupligs 
$A_k^{ij}$ (common trilinear soft breaking term for each 
trilinear term
at the Planck scale)

4. All soft breaking parameters are real.

\noindent As one can see these assumptions greatly reduce the 
number of 
independent free parameters of the theory. However one has to
stress that these are just assumptions, with no solid basis 
of origin. The strongest argument in favor of these assumptions
is that if one takes a supergravity theory in which SUSY
is broken in a hidden sector and SUSY breaking is 
communicated to the visible sector by gravity then one gets
flavor independent mass terms, real universal 
$A$-terms at the Planck scale and real gaugino masses, provided
one assumes that the K\"ahler potential of the supergravity
theory is minimal. 

The argument for gaugino unification is the following. It is
experimentally indicated that gauge couplings do unify in the 
MSSM 
\cite{Amaldi}. However
the 1-loop RGE for the gaugino masses is given by \cite{Diego}
\begin{equation}
\frac{d}{dt} \left( \frac{\alpha_i}{M_i} \right) =0, \; \; 
i=1,2,3 \; \; t=\log \left( \frac{\Lambda}{M_{GUT}} \right). 
\end{equation}

Here $\alpha_i=g_i^2/4\pi$, $g_i$ are the gauge couplings and
$M_i$ the gaugino masses. The ratios of gauge couplings to
gaugino masses are scale invariant. Thus if gauge couplings 
unify so must the gaugino masses.

If one accepts these arguments then the independent soft breaking
terms are $A_0, m_0, B$ and $M_{1/2}$ (at the Planck scale), 
and the
soft breaking Lagrangian  at the Planck-scale is given by

\begin{eqnarray}
\label{softpot}
& & -{\cal L}_{soft}|_{M_{P}}= 
m_0^2\sum_{i=Q_{i},\bar{U}_{i},...}
 |\phi_i|^2+ \left[
M_{1/2}\sum_{i=1,2,3} \lambda_i \lambda_i - B\mu H_1H_2+ \right.
\nonumber \\
& & +\left.
A_0(\sum_{ij} \lambda_u^{ij} Q^iH_2\bar{U}^j+
\sum_{ij} \lambda_d^{ij}Q^iH_1\bar{D}^j
+\sum_{ij} \lambda_e^{ij}L^iH_1\bar{E}^j) + h. c. \right],
\end{eqnarray}
and the Lagrangian at the weak scale can be obtained by running
down these universal parameters from the Planck-scale\footnote{
Usually the Planck-scale and the GUT-scale are not distinguished
 and it is common to assume that \ref{softpot} is still valid 
at the GUT-scale $\approx 10^{16}$ GeV.} to the weak scale.
This procedure will yield sufficiently degenerate squark and
slepton masses, and if the soft breaking terms are real at
the Planck scale then they will not obtain imaginary parts
at the weak scale either. Therefore the above assumptions 
do satisfy the phenomenological constraints and at the same
time they greatly increase the predictive power of the theory.
Often these assumptions about the soft breaking terms are 
assumed to be part of the definition of the MSSM. However
one can not overemphasize the fact that these assumptions are 
ad hoc and need not necessarily be satisfied.

In the remainder of this Section we will discuss the breaking
of SU(2)$\otimes $U(1). The Higgs potential without soft 
breaking  terms is given by
\begin{equation}
\label{higgs1}
V_{SUSY} (H_1,H_2)=\mu^2 (|H_1|^2+|H_2|^2) +
\frac{g^2}{2} (H_1^{\dagger} \vec{\tau}H_1+H_2^{\dagger}
\vec{\tau}H_2)^2+\frac{g'^2}{2} (H_1^{\dagger}H_1-H_2^{\dagger}
H_2)^2.
\end{equation}
The minimum of this potential is at $\langle H_1 \rangle =
\langle H_2 \rangle = 0$, thus we need to incorporate the
soft breaking terms to get electroweak breaking. The full Higgs
potential at the Planck (GUT) scale is
\begin{eqnarray}
\label{higgs2}
V (H_1,H_2)|_{GUT}&=& (\mu^2+m_0^2) (|H_1|^2+|H_2|^2) -B\mu 
(H_1H_2+h.c.)+\nonumber \\ &+&
\frac{g^2}{2} (H_1^{\dagger} \vec{\tau}H_1+H_2^{\dagger}
\vec{\tau}H_2)^2+\frac{g'^2}{2} (H_1^{\dagger}H_1-H_2^{\dagger}
H_2)^2.
\end{eqnarray}
This potential still does not break SU(2)$\otimes $U(1). This
can be seen in the following way: in order to have a nontrivial
minimum of the Higgs potential
\begin{equation}
\label{higgs3}
m_{H_{1}}^2 |H_1|^2+m_{H_{2}}^2|H_2|^2 -m_{12}^2 (H_1H_2+h.c.)+
\frac{g^2}{2} (H_1^{\dagger} \vec{\tau}H_1+H_2^{\dagger}
\vec{\tau}H_2)^2+\frac{g'^2}{2} (H_1^{\dagger}H_1-H_2^{\dagger}
H_2)^2
\end{equation}
the quadratic coefficients have to fulfill the following 
inequalities:
\begin{eqnarray}
\label{ineq}
& & m_{H_{1}}^2+m_{H_{2}}^2 > 2|m_{12}^2| \nonumber \\
& & |m_{12}^2|^2>m_{H_{1}}^2m_{H_{2}}^2.
\end{eqnarray}
The first inequality is required so that the potential remains
bounded from below for the equal field direction $H_1=H_2$, 
while the second is required so that the quadratic piece 
has a negative part enabling a nontrivial minimum.  We can see
that the potential in eq. \ref{higgs2} can not fulfill both
inequalities at the same time thus electroweak symmetry is
not broken at the tree level. However radiative corrections 
can change
this situation. To calculate these radiative effects one needs 
to evaluate 
the one loop effective potential:
\begin{equation}
V_{1-loop}=V_{tree} (\Lambda ) +\Delta V_1 (\Lambda ),
\end{equation}
where $V_{tree}$ is the tree level superpotential with running 
parameters
evaluated at a scale $\Lambda $ and $\Delta V_1$ is the 
contribution
of one loop diagrams to the effective potential evaluated by 
the method
of Coleman and Weinberg. The running of the parameters in the 
tree
level potential is generated by the one loop RGE's. 
$V_{tree}+\Delta V_1$
is $\Lambda $ independent up to one loop order. If we choose 
the scale
$\Lambda $ to be close to the scale of the masses of the
particles of the theory (in our case $\Lambda \simeq M_{weak}$) 
$\Delta V_1$ will not contain large logarithms, thus the 
leading one loop
effects will arise due to the running of the parameters of 
the tree level 
potential between the Planck and the weak scale.  
To estimate
the running effects on the Higgs parameters we neglect all
Yukawa couplings with the exception of the top Yukawa coupling
(this is the only large Yukawa coupling so it is reasonable
to assume that the largest effects will be caused by it).
Then the RGE's for the
soft breaking mass terms of the scalars participating in the
top Yukawa coupling of the superpotential are 
\cite{Diego}:
\begin{eqnarray}
\label{RGE1}
& & \frac{dm_{H_{2}}^2}{dt}=\frac{3}{5}g_1^2M_1^2+3g_2^2M_2^2
-3\lambda_t^2(m^2+A_t^2) \nonumber \\
& & \frac{dm_{\tilde{t}}^2}{dt}=\frac{16}{15}g_1^2M_1^2+
\frac{16}{3}g_3^2M_3^2-2\lambda_t^2(m^2+A_t^2) \nonumber \\
& & \frac{m_{\tilde{Q}_{3}}^2}{dt}=\frac{1}{15}g_1^2M_1^2+
3g_2^2M_2^2+\frac{16}{3}g_3^2M_3^2-\lambda_t^2(m^2+A_t^2),
\end{eqnarray}
where $m^2= m_{H_{2}}^2+m_{\tilde{t}}^2+m_{\tilde{Q}_{3}}^2$,
$t=\frac{1}{16\pi^2} \log \frac{M_{GUT}^2}{\Lambda^2}$, 
$\Lambda$ is the
energy scale, $g_i$ are the gauge couplings, $A_t$ the
soft breaking trilinear parameter corresponding to the
top Yukawa coupling, $M_i$ the gaugino masses and $\lambda_t$
the top Yukawa coupling.

One can see that the contributions of the gauge and Yukawa loops
are independent of each other and the contributions of the
gauge loops are independent of the soft breaking masses
$m_i^2$. Thus one can solve eq. \ref{RGE1} by setting the gauge 
couplings to zero and at the end add the gauge contribution
to the resulting solution. Therefore one has to solve the 
following equation:
\begin{equation}
\label{RGE}
\frac{d}{dt} \left( \begin{array}{c} m_{H_{2}}^2 \\ 
m_{\tilde{t}}^2 \\ m_{\tilde{Q}_{3}}^2 \end{array} \right)
= -\lambda_t^2 \left( \begin{array}{ccc} 3 & 3& 3 \\ 2&2&2\\
1&1&1 \end{array} \right) \left( \begin{array}{c} m_{H_{2}}^2 \\ 
m_{\tilde{t}}^2 \\ m_{\tilde{Q}_{3}}^2 \end{array} \right) 
-\lambda_t^2 A_t^2 \left( \begin{array}{c} 3\\ 
2 \\1 \end{array} \right).
\end{equation}
This differential equation can be solved easily if one neglects
 the running of of $\lambda_t$ and $A_t$. The solution 
corresponding to the universal boundary condition
at $t=0$ $(\Lambda =M_{GUT})$ $m_{H_{2}}^2=m_{\tilde{t}}^2=
m_{\tilde{Q}_{3}}^2=m_0^2$ in the limit $t\rightarrow \infty$
is given by
\begin{eqnarray}
\label{sol}
& & m_{H_{2}}^2=-\frac{1}{2}m_0^2 \nonumber \\
& & m_{\tilde{t}}^2=0 \nonumber \\
& & m_{\tilde{Q_{3}}}^2=\frac{1}{2}m_0^2.
\end{eqnarray}
Thus we can see that the radiative corrections 
due to the top Yukawa coupling want to reverse the sign
of the soft breaking mass parameter of the up-type Higgs,
which is enough to satisfy the conditions for electroweak
breaking of eq. \ref{ineq} at the weak scale. The gauge loops 
will yield additional positive contributions proportional
to $M_{1/2}^2$, and the solution to \ref{RGE} is
more complicated if one takes the running of $\lambda_t$ and
$A_t$ into account. However the most important 
feature of the solution in \ref{sol} is unchanged: 
appropriate choices of the input parameters $M_{1/2},m_0, A_0$ 
and $\lambda_t$ will drive the soft breaking mass parameter
of the up-type Higgs (and only of the up-type Higgs) negative
which will result in the breaking of electroweak symmetry.
This mechanism is
called radiative electroweak breaking. 

Thus as we have seen loop corrections usually modify the 
Higgs potential such that at the weak scale SU(2)$\otimes $
U(1) is spontaneously broken. However it is not enough 
to require that the symmetry is broken, it has
to reproduce the correct SM minimum. The Higgs potential
at the weak scale can be written as

\begin{eqnarray}
\label{higgs4}
&&(m_{H_{1}}^2+\mu^2) |H_1|^2+(m_{H_{2}}^2+\mu^2)|H_2|^2 
-B\mu (H_1H_2+h.c.)+\nonumber \\ &&
\frac{g^2}{2} (H_1^{\dagger} \vec{\tau}H_1+H_2^{\dagger}
\vec{\tau}H_2)^2+\frac{g'^2}{2} (H_1^{\dagger}H_1-H_2^{\dagger}
H_2)^2.
\end{eqnarray}
The VEV's of the Higgs doublets are 
\begin{equation}
\langle H_1 \rangle =\left( \begin{array}{c} 0 \\v_1 \end{array}
\right), \; \langle H_2 \rangle =\left( \begin{array}{c} v_2 
\\0 \end{array} \right),
\end{equation}
and we define $\tan \beta =v_2/v_1$, $v^2=v_1^2+v_2^2$.
Minimizing the Higgs potential we find that to fix the W,Z 
masses at their experimental values it is necessary that 
\cite{Diego}
\begin{eqnarray}
\label{ewbr}
& & \mu^2=\frac{m_{H_{1}}^2-m_{H_{2}}^2\tan^2 
\beta}{\tan^2 \beta -1}-\frac{1}{2}M_Z^2 \nonumber \\
& & B=\frac{(m_{H_{1}}^2+m_{H_{2}}^2+2\mu^2)\sin 2\beta}{2
\mu},
\end{eqnarray}
where all parameters are to be evaluated at the weak scale.

With this we are now able to determine the free parameters
of the MSSM. In the soft breaking sector we had $m_0,M_{1/2},
A_0$ and $B$. In the Higgs sector we have $\mu$  and 
$\tan \beta$, and since the top mass is experimentally not well
measured and $\lambda_t$ tends to run to an IR fixed point
at $M_Z$, $\lambda_t (M_G)$ is basically an unknown parameter
of the theory as well. However from eq. \ref{ewbr} $\mu^2$ and
$B$ are determined (but not the sign of $\mu$). Therefore 
the MSSM with R-parity, universal soft breaking parameters
and radiative electroweak breaking is determined by 5+1 
parameters: $m_0, M_{1/2},A_0,\tan \beta, \lambda_t$ and the
sign of $\mu$.

\mysection{Sparticle masses}

In this section we present the possible mixings between 
the superpartner fields and list the tree level mass matrices
\cite{Haber,Diego,Ellis}.

\subsection{Sfermions}

In principle one must diagonalize 6 by 6 matrices 
corresponding to the mixing of the L and R scalars of the
3 generations. To simplify this we neglect intergenerational 
mixings and take only L-R mixing into account. 

\subsubsection{Squarks}

The mass matrices in the L-R basis are for each generation of 
up-type scalars is 
\begin{equation} 
M^2_{\tilde{u}_{L,R}}=\left( \begin{array}{cc} 
m_{\tilde{Q}}^2+m_u^2+(\frac{1}{2}-\frac{2}{3}\sin^2 \theta_W)D
& m_u(A_u-\mu \cot \beta ) \\  m_u(A_u-\mu \cot \beta ) &
m_{\tilde{u}}^2+m_u^2+\frac{2}{3}\sin^2 \theta_W D \end{array}
\right),
\end{equation}
where the mass parameters with a tilde refer to the soft breaking
squark mass parameters while the mass parameters without tilde 
are the
ususal quark masses, $D=M_Z^2\cos 2\beta$.
 
The down-type mass matrix is 
\begin{equation} 
M^2_{\tilde{d}_{L,R}}=\left( \begin{array}{cc} 
m_{\tilde{Q}}^2+m_d^2-(-\frac{1}{2}-\frac{1}{3}\sin^2 \theta_W)D
& m_d(A_d-\mu \tan \beta ) \\  m_d(A_d-\mu \tan \beta ) &
m_{\tilde{d}}^2+m_d^2-\frac{1}{3}\sin^2 \theta_W D \end{array}
\right).
\end{equation}

The only source of intergenerational mixing is the 
superpotential, thus in the more general case the diagonal 
elements $m_u^2$ and $m_d^2$ must be exchanged to
$v_{2,1}^2 (\lambda_{u,d}^{\dagger}\lambda_{u,d})_{ij}$,
where $ij$ are generation indices. However since the CKM 
mixing is small and the soft breaking mass terms are large
compared to the quark masses, these effects are usually 
negligible.

\subsubsection{Sleptons}

In the same notation the sneutrino masses are:

\begin{equation}
M_{\tilde{\nu}}^2=M_{\tilde{L}}^2+\frac{1}{2}D
\end{equation}
while for $\tilde{e}, \tilde{\mu}, \tilde{\tau}$ the mass 
matrices are
\begin{equation} 
M^2_{\tilde{e}_{L,R}}=\left( \begin{array}{cc} 
m_{\tilde{L}}^2+m_e^2-(\frac{1}{2}-\sin^2 \theta_W)D
& m_e(A_e-\mu \tan \beta ) \\  m_e(A_e-\mu \tan \beta ) &
m_{\tilde{e}}^2+m_e^2-\sin^2 \theta_W D \end{array}
\right).
\end{equation}

\subsection{The scalar Higgs sector}

We use the notation
\begin{equation}
H_1= \left( \begin{array}{c} h_1^0 \\ h_1^- \end{array} 
\right), \; \; H_2= \left( \begin{array}{c} h_2^+ \\ h_2^0 
\end{array} 
\right). 
\end{equation}
The tree level masses are calculated from the mass matrices

\begin{eqnarray}
&&\frac{1}{2} \frac{\partial^2 V_{tree}}{\partial (Im h_i^0) 
\partial (Im h_j^0)}= \frac{1}{2} M_A^2 \sin 2\beta 
\left( \begin{array}{cc} \tan \beta & 1 \\ 1 & \cot \beta 
\end{array} \right) \nonumber \\
&&\frac{1}{2} \frac{\partial^2 V_{tree}}{\partial (Re h_i^0) 
\partial (Re h_j^0)}= \frac{1}{2} M_A^2 \sin 2\beta 
\left( \begin{array}{cc} \tan \beta & -1 \\ -1 & \cot \beta 
\end{array} \right)+\frac{1}{2} M_Z^2 \sin 2\beta 
\left( \begin{array}{cc} \cot \beta & -1 \\ -1 & \tan \beta 
\end{array} \right)
\nonumber \\
&&\frac{\partial^2 V_{tree}}{\partial h_i^- 
\partial h_j^+}= \frac{1}{2} M_{H^{\pm}}^2 \sin 2\beta 
\left( \begin{array}{cc} \tan \beta & 1 \\ 1 & \cot \beta 
\end{array} \right),
\end{eqnarray}
where $M_A^2=m_{H_{1}}^2+m_{H_{2}}^2+2\mu^2$, $M_{H^{\pm}}=
M_W^2+M_A^2$, $i,j=1,2$.

The first mass matrix has eigenvalues 0 (GB eaten by the Z) and
$M_A^2$ (CP odd scalar). The second matrix gives the masses
for the light and heavy Higgs bosons:
\begin{equation}
M_{H,h}^2=\frac{1}{2}\left[ 
(M_A^2+M_Z^2)\pm \sqrt{(M_A^2+M_Z^2)^2-
4M_A^2M_Z^2\cos^22\beta}\right].
\end{equation}
The third matrix has eigenvalues 0 (charged GB's eaten by 
$W^{\pm}$) and $M_{H^{\pm}}^2$ (charged scalars).

It is important to mention that for some of the Higgs masses
the 1-loop corrections can be significant. For example from the
above formula one would get that $m_h \leq M_Z$, while
including 1-loop corrections the corresponding bound will
be modified to $m_h \leq 150$ GeV \cite{Ellis}. 

\subsection{Charginos}

Charginos are mixtures of the charged Higgsinos  and the
charged gauginos ($\tilde{W}_{1,2}$). The mass matrix is 
given by 
($\lambda^{\pm}=(\tilde{W}_2\pm i\tilde{W}_1)/\sqrt{2}$):

\begin{equation}
\left( \begin{array}{cccc} \lambda^+ & \tilde{h}_2^+ &
\lambda^- & \tilde{h}_1^- \end{array} \right)
\left( \begin{array}{cccc} 0&0& M_2&-g_2v_1 \\
0&0&g_2v_2&-\mu \\ M_2&g_2v_2&0&0\\-g_2v_1&-\mu&0&0 \end{array}
\right) \left( \begin{array}{c} \lambda^+ \\ \tilde{h}_2^+ \\
\lambda^- \\ \tilde{h}_1^- \end{array} \right).
\end{equation}
The eigenvalues are
\begin{equation}
M_{\tilde{C}_{1,2}}=\frac{1}{2} \left[(M_2^2+\mu^2+2M_W^2)
\pm \sqrt{(M_2^2+\mu^2+2M_W^2)^2-4(M_2\mu-M_W^2\sin 2\beta)^2}
\right].
\end{equation}

\subsection{Neutralinos}

Neutralinos are the mixture of neutral Higgsinos and the
neutral gauginos ($\tilde{B},\tilde{W}_3$). The mass matrix is
given by
\begin{equation}
\left( \begin{array}{cccc} i\tilde{B} & i\tilde{W}_3 &
\tilde{h}_1^0 & \tilde{h}_2^0 \end{array} \right)
\frac{1}{2} \left( \begin{array}{cccc} -M_1&0&g'v_1/\sqrt{2}&
-g'v_2/\sqrt{2} \\ 0&-M_2&-g_2v_1/\sqrt{2}&g_2v_2/\sqrt{2} \\
g'v_1/\sqrt{2} & -g_2v_1/\sqrt{2} & 0 & \mu \\
-g'v_2/\sqrt{2}&g_2v_2/\sqrt{2}&\mu &0 \end{array} \right)
\left( \begin{array}{c}  i\tilde{B} \\ i\tilde{W}_3 \\
\tilde{h}_1^0 \\ \tilde{h}_2^0 \end{array} \right).
\end{equation}

\mysection{Conclusions}

We have systematically built up the minimal supersymmetric 
standard model. We have started from the particle content of
the theory, discussed the superpotential and R-parity. Then
the question of SUSY breaking and electroweak symmetry 
breaking have been examined. Finally we listed the physical 
particles of the theory, together with the tree level mass 
matrices.

\section*{Acknowledgements}

I have learned large portions of the material presented here
from discussions with Diego Casta\~no, Daniel Freedman and
Lisa Randall and from a series of lectures on this subject
at TASI '95 presented by Jonathan Bagger and Xerxes Tata.
I am especially grateful to Daniel Freedman for carefully
reading this manuscript and for his encouragement to publish 
this paper originally written as a term paper for his course
at MIT on supersymmetry.


\begin{thebibliography}{99}

\bibitem{Haber}
H. Haber, SCIPP-92-33, hep-ph/9306207.

\bibitem{Nilles}
H-P. Nilles, TUM-HEP-230-95, hep-ph/9511313.

\bibitem{Baer}
H. Baer et. al., FSU-HEP-950401, hep-ph/9503479.

\bibitem{Bagger}
J. Bagger, JHU-TIPAC-96008, hep-ph/9604232.

\bibitem{Diego}
D.J. Castano, E.J. Piard and P. Ramond, Phys. Rev. {\bf D49}
(1994), 4882.

\bibitem{Martin}
M. Drees and S. Martin, MADPH-95-879, UM-TH-95-02,hep-ph/9504324.

\bibitem{Ibanez}
L.E. Ibanez and C. Lopez, Nucl. Phys. {\bf B233} (1984), 511, \\
L.E. Ibanez, C. Lopez and C. Munoz, Nucl. Phys. {\bf B256} 
(1985), 218.

\bibitem{WB}
J. Wess and J. Bagger, Supersymmetry and Supergravity (Princeton
Univ. Press. 1992).

\bibitem{Ross}
G.G. Ross, Grand Unified Theories (Addison Wesley 1984).

\bibitem{Exp}
X. Tata, UH-511-833-95, hep-ph/9510287.

\bibitem{DimGeor}
S. Dimopoulos and H. Georgi, Nucl. Phys. {\bf B193} (1981), 150.

\bibitem{GG}
L. Girardello and M.T. Grisaru, Nucl. Phys. {\bf B194} (1984),
419.

\bibitem{Amaldi}
U. Amaldi et. al., Phys. Rev. {\bf D36} (1987) 1385.

\bibitem{Ellis}
J. Ellis, G. Ridolfi, F. Zwirner, Phys. Lett. {\bf B257} (1991),
83, \\
Phys. Lett. {\bf B262} (1991), 477.

\end{thebibliography}
\end{document}